\documentclass[]{frank}              	

\usepackage{amsmath}
\usepackage{amscd, amsthm, amsfonts, amssymb}
\usepackage{mathrsfs, bbm, cancel, cite, times}
\usepackage{marvosym, enumerate, enumitem, xfrac, verbatim}
\usepackage[all]{xy}
\usepackage[normalem]{ulem}
\usepackage{dsfont}
\usepackage{graphicx}
\usepackage{hyperref, color}
\definecolor{darkblue}{rgb}{0,0,1}
\hypersetup{pdfpagemode=UseNone, colorlinks=true, breaklinks=true, linkcolor=darkblue, menucolor=darkblue, urlcolor=darkblue, citecolor=black}

\newtheorem{prop}{Proposition}

\theoremstyle{remark}

\theoremstyle{definition}

\newtheorem{rem}[prop]{Remark}

\newcommand{\q}{{\rm q}}

\newcommand{\RR}{\mathbbm R}

\newcommand{\CC}{\mathbbm C}

\allowdisplaybreaks 

\begin{document}
\sloppy \raggedbottom

\title{Eleven-dimensional symmetric supergravity backgrounds, their geometric superalgebras, and a common reduction\thanks{
		Extended notes of a talk given at the Mini-Symposium {\em Algebraic Methods in Quantum Field Theory} as part of the {\em International Conference
			``Mathematics Days in Sofia'' July 7-10, 2014,}
	}}
	\runningheads{$D=11$ symmetric supergravity backgrounds}{F.~Klinker}

\begin{start}
	\author{Frank Klinker}{1}
	
	\address{Faculty of Mathematics, TU Dortmund University, 44221 Dortmund, Germany\\[1ex]
		\href{mailto:frank.klinker@math.tu-dortmund.de}{frank.klinker@math.tu-dortmund.de}}{1}
	
	\received{ }	
	
	\begin{Abstract}
		We present two different families of eleven-dimensional manifolds that admit non-restricted extensions of the isometry algebras to geometric superalgebras. Both families admit points for which the superalgebra extends to a super Lie algebra; on the one hand, a family of $N=1$, $\nu={}^3\!/\!_4$ supergravity backgrounds and, on the other hand, a family of $N=1$, $\nu=1$ supergravity background. Furthermore, both families admit a point that can be identified with an $N=4$, $\nu={}^1\!/\!_2$ six-dimensional supergravity background.
	\end{Abstract}
\PACS {04.65.+e, 02.40.Hw, 12.60.Jv}
\end{start}	


\section{The setup}
\subsection{CW-spaces}

In this text we consider CW-spaces and ask about the conditions such that they can be considered as backgrounds in supergravity. In this context CW-spaces have been discussed in \cite{CKG,FigPap,GauntHull,Fig03,MeFig04,Hustler}, for example.
CW-spaces are Lorentzian solvable symmetric spaces that has been characterized in the early 1970's by M.~Cahen and N.~Wallach, see \cite{CW}. There is a one-to-one correspondence between $D=n+2$-dimensional CW-spaces and triples $(V,B,\langle\cdot,\cdot\rangle)$ of an $n$-dimensional euclidean vector space $V$, a symmetric map $B\in End(V)$, and an extension $\langle\cdot,\cdot\rangle$ of the euclidean product on $V$ to a block-diagonal Lorentzian metric on $W=\RR^2\oplus V$. We fix a null basis $\{e_+,e_-\}$ of the $\RR^2$-factor.  The manifold structure of the CW-space is infinitesimally defined by a Lie algebra structure on $V^*\oplus W$ of which the non-vanishing brackets are given by
\begin{align}
 [v^*,e_-]= Bv,\ 
 [v^*,w]= -\langle Bv,w\rangle e_+,\  
 [e_-,w]= w^*\,.
\end{align}
Here, $*:V\to V^*$ is defined by $\langle\cdot,\cdot\rangle$. In particular, two such spaces $M_{B},M_{\tilde B}$ are isometric if and only if there exists an orthogonal map $A$ and a positive scalar $\beta$ such that $\tilde B =\beta A BA^t$, see \cite{CW,Wu}. The CW-space $M_B$ that is defined by such triple is indecomposable if and only if $B$ is non-degenerate. This can for example be seen by taking a look at the explicit form of the metric (\ref{metric}).

\subsection{The metric and the Killing vector fields}
Using the exponential map $\mu(x)=\exp(x^+e_+)\exp(x^-e_-)\exp(\sum_ix^ie_i)$ we get 
\begin{equation*}
\begin{aligned}
\mu^{-1}d\mu=\ & \sum_ix^-dx^-\otimes e_i^* \\ &+dx^-\otimes e_-+\sum_idx^i\otimes e_i+(dx^+-\tfrac{1}{2}\sum_{ij}B_{ij}x^ix^jdx^-)\otimes e_+
\end{aligned}
\end{equation*}
such that the local metric of the $D=n+2$-dimensional CW-space $M_B$ is 
\begin{equation}\label{metric}
g=2dx^+dx^--\sum_{ij}B_{ij}x^ix^j(dx^-)^2+\sum_i(dx^i)^2\,.
\end{equation}  
We can and will in the following consider an orthonormal basis $\{e_i\}$ of $V$ such that the symmetric map $B$ is diagonal, namely $B={\rm diag}(\lambda_1^2,\ldots,\lambda_n^2)$. If 
$\lambda_1^2=\cdots=\lambda_{r_1}^2<\lambda_{r_1+1}^2=\ldots=\lambda_{r_2}^2<\cdots<\lambda_{r_{\hat n-1}+1}^2=\cdots=\lambda_n^2$ we write 
$\{1,\ldots,n\}=\bigcup_{\alpha=1}^{\hat n}$ with 
$I_\alpha=\{r_{\alpha-1},\ldots,r_\alpha\}$ and $r_0=1,r_{\hat n}=n$.

The Killing vector fields of this metric are given by 
\begin{equation}\label{killing}
\begin{aligned}
K_{(+)}  &=-\partial_+,\quad K_{(-)}=-\partial_- \\
K_{(i)}  &= \cos(\lambda_ix^-)\partial_i+\lambda_i\cos(x^-)x^i\partial_+\,,\quad  i=1,\ldots,n, \\
K_{(i^*)}&= -\lambda_i\sin(\lambda_ix^-)\partial_i+\lambda_i^2\cos(x^-)x^i\partial_+\,,\quad i^*=1,\ldots,n,\\
K_{(ij)} &= x^j\partial_i-x^i\partial_j\,,\quad (ij)\in I_1^2\cup\ldots\cup I_{\hat n}^2\,.
\end{aligned}
\end{equation}
We denote the Lie algebra that is spanned by the Killing vector fields by $\mathcal{K}(0)$, see also \cite{CKG,FigPap}. 

\subsection{Connections on the spinor bundle}
We consider the irreducible spin bundle $S$ over $M_B$. 
We denote the images of the basis $\{e_+,e_-,e_i\}$ under the spin representations by $\{\Gamma_+,\Gamma_-,\Gamma_i\}$.
They obey the usual Clifford relation $\Gamma_A\Gamma_B+\Gamma_B\Gamma_A=-2g_{AB}$. 
The spin bundle splits into two subbundles $S=S_-\oplus S_+$ where the first and second summand are the $-1$- and $+1$ eigenspaces of $\sigma:=\frac{1}{2}[\Gamma_+\Gamma_-]$, respectively.
The two projections on the subbundles are given by $\sigma_{\pm}=-\frac{1}{2}\Gamma_\mp\Gamma_\pm$. 
Due to $\Gamma_+^2=0$ we have $\Gamma_+:S_+\to S_-$ and $S_+={\rm ker}\Gamma_+$. We denote the components of a section $\xi\in\cancel{S}$ with respect to the above decomposition by $\xi=\xi_1+\xi_2$. 

The Levi-Civita connection on $M_B$ induces a connection on the spinor bundle. It is given by 
\begin{equation}\begin{aligned}
\nabla_+\xi &=\partial_+\xi,\quad \nabla_i\xi=\partial_i\xi\\
\nabla_-\xi &=\partial_-\xi-\frac{1}{2}\sum_ix^i\Gamma_+B(e_i)\xi_2\,.
\end{aligned}
\end{equation} 
From \cite{Klinker2014a} we know that any connection $D$ on $S$ that is compatible with the symmetric structure of $M_B$ is described by a pair of elements $(c,d)$ of the Clifford algebra of $V$. It is given by
\begin{equation}\label{conn-gen}
\begin{aligned}
D_+\xi_1 &=\nabla_+\xi_1\,,											& D_+\xi_2&=\nabla_+\xi_2\,,\\
D_-\xi_1 &=\nabla_-\xi_1+ c\xi_1\,, 								& D_-\xi_2&=\nabla_-\xi_2+ d\xi_2\,,\\
D_i\xi_1 &=\nabla_i\xi_1-\frac{1}{2}\Gamma_+ s_{c,d}(e_i)\xi_2\,, 	& D_i\xi_2&=\nabla_i\xi_2\,.
\end{aligned}
\end{equation}
with $s_{c,d}(v)$ given by
\begin{equation}
s_{c,d}(e_i)=c\Gamma_i-\Gamma_id\,.
\end{equation}
\begin{rem}
If the spinor bundle is reducible and of the form $S\otimes\CC^N$ the result on the form of the connection remains true but with the parameters $c,d$ taking their values in the tensor product of the Clifford algebra and $\mathfrak{gl}_N\CC$.
\end{rem}
The parallel spinors with respect to this connection are given by $\xi=(\xi_1,\xi_2)$ with
\begin{equation}
\begin{aligned}
\xi_1 &= \exp(x^-c)\xi_1^0  -\frac{1}{2}\sum_i x^i\Gamma_+s_{c,d}(e_i)\exp(x^-d)\xi_2^0\,,\\
\xi_2 &= \exp(x^-d)\xi_2^0\,,
\end{aligned}
\end{equation}
where $\xi^0=(\xi_1^0,\xi_2^0)$ is a constant spinor subject to the condition
\begin{equation}
(q_{c,d}(e_i)+B(e_i))\xi_2=0
\end{equation}
with
\begin{equation}
q_{c,d}(e_i)=s_{c,d}\circ s_{c,d}(e_i)= c^2\Gamma_i+\Gamma_i d^2-2c\Gamma_i d\,.
\end{equation}

\section{Geometric superalgebras}

A geometric superalgebra of a CW-space $M_B$ is an extension of $\mathcal{K}_0$ to a superalgebra $\mathcal{K}_0\oplus\mathcal{K}_1$ with the following properties:
\begin{enumerate}
\item $\mathcal{K}_1$ is a subset of the space of the sections that are parallel with respect to a connection $D$ on a spinor bundle $S$.
\item There exist a linear map $\mathcal{L}:\mathcal{K}_0\to End(\mathcal{K}_1)$, such that $[\mathcal{L}_X\mathcal{L}_Y]=\mathcal{L}_{[X,Y]}$.
\item There exist a bilinear symmetric map $\{\cdot,\cdot\}:\mathcal{K}_1\times\mathcal{K}_1\to \mathcal{K}_0$ such that $2\{\mathcal{L}_X\xi,\xi\}=[X,\{\xi,\xi\}]$.
\end{enumerate}
A geometric superalgebra is called irreducible if $S$ is as a Clifford module. In case that the spinor bundle is reducible with $S\otimes\CC^N$ we call the superalgebra $N$-extended; sometimes we write $N=1$ instead of irreducible.  
A geometric superalgebra is called non-restricted if the space $\mathcal{K}_1$ contains all parallel spinors, otherwise it is called restricted. We denote by $\nu$ the quotient of ${\rm dim}\mathcal{K}_1$ and ${\rm dim} S$. Then a superalgebra extension of a CW-space with flat connection is non-restricted if and only if $\nu=1$. Nevertheless, a superalgebra extension with $\nu<1$ can be restricted or a non-restricted.

A geometric superalgebra is called supersymmetry algebra if the extension actually is a super Lie algebra, i.e.\ if in addition
\begin{enumerate}\addtocounter{enumi}{3}
	\item\label{item:susy} $\displaystyle\mathcal{L}_{\{\xi,\xi\}}\xi=0$ is fulfilled for all $\xi\in\mathcal{K}_1$.
\end{enumerate}
In this situation we also say that the underlying space admits geometric supersymmetry.

In our situation $\mathcal{L}$ is the spinorial Lie derivative, see \cite{Kosman}. It is properly defined for Killing vector fields $X$ and given by
\begin{equation}
	\mathcal{L}_X\xi=\nabla_X\xi-\Gamma(\nabla X)\xi\,.
\end{equation}
In particular, for this map the second part of 2.\ is satisfied. 

From now on we restrict to $D=11$ and $n=9$.

\subsection{A family of $N=1$, $\nu=1$ geometric superalgebras}

We consider connection (\ref{conn-gen}) with 
\begin{equation}
c=\alpha\Gamma_{123}, d=\beta\Gamma_{123}\,.
\end{equation}
These Clifford elements obey
\begin{equation}
q_{c,d}(e_i)=
\begin{cases}
	(\alpha-\beta)^2\Gamma_i&\text{ for }i\in\{1,2,3\}\\
	(\alpha+\beta)^2\Gamma_i&\text{ for }i\in\{4,\ldots,9\}
\end{cases}
\end{equation}
Therefore, $(c,d)$ yields a flat connection on the CW-space $M_B$ that is defined by \begin{equation}\label{metric-1}
B=-{\rm diag}\left((\alpha-\beta)^2\mathbbm{1}_3,(\alpha+\beta)^2\mathbbm{1}_6\right)\,.
\end{equation} 
It is indecomposable if and only if $\alpha\neq\pm\beta$. 
We write $\lambda_j^2=-(\alpha-\beta)^2$ for $j\in \{1,2,3\}$ and $\lambda_j^2=-(\alpha+\beta)^2$ for $j\in \{4,\ldots,9\}$ and set $\lambda_j=-i(\alpha\pm\beta)$.
 
If we consider
\begin{equation} 
\mathcal{K}_1=\left\{\xi\in\cancel S\left|
	\begin{array}{l@{\,}l}
		\xi=\xi(\xi_1^0,\xi_2^0)= &  \exp(\alpha x^-\Gamma_{123})\xi_1^0 \\
		& +\Big(1-\frac{1}{2}\sum_i\Gamma_+x^i(\alpha\Gamma_{123}\Gamma_i-\beta\Gamma_i\Gamma_{123})\Big)\\
		& \quad\cdot\exp(\beta x^-\Gamma_{123})\xi_2^0  \\[0.5ex]
		\multicolumn{2}{l}{\xi_1^0,\xi_2^0\text{ constant},\ \sigma_-\xi_2^0=\sigma_+\xi_1^0=0}
	\end{array}
\right.\right\}
\end{equation}
In particular we have $\dim\mathcal{K}_1=\dim S$ such that $\nu=1$. 

For $\xi=\xi(\xi_1^0,\xi_2^0)\in\mathcal{K}_1$ we have $\mathcal{L}_X\xi\in\mathcal{K}_1$; more precisely
\begin{equation}\label{L-rel}
\begin{aligned}
\mathcal{L}_{K_{(+)}}\xi(\xi_1^0,\xi_2^0)   &=\xi(0,0)\,,\\
\mathcal{L}_{K_{(-)}}\xi(\xi_1^0,\xi_2^0)   &=\xi(-c\xi_1^0,-d\xi_2^0)\,,\\
\mathcal{L}_{K_{(i)}}\xi(\xi_1^0,\xi_2^0)   &=\xi(\tfrac{1}{2}\Gamma_+s_{c,d}\xi_2^0,0)\,,\\
\mathcal{L}_{K_{(i^*)}}\xi(\xi_1^0,\xi_2^0) &=\xi(-\tfrac{1}{2}\Gamma_+B(e_i)\xi_2^0,0)\,,\\
\mathcal{L}_{K_{(ij)}}\xi(\xi_1^0,\xi_2^0)&=\xi(\tfrac{1}{2}\Gamma_{ij}\xi_1^0,\tfrac{1}{2}\Gamma_{ij}\xi_2^0)\,.
\end{aligned}
\end{equation}
To complete the geometric superalgebra wee need the map $\{\cdot,\cdot\}:\mathcal{K}_1\times\mathcal{K}_1\to\mathcal{K}_0$. 
We will write down $\{\xi,\xi\}$ for $\xi=\xi(\xi_1^0,\xi_2^0)$ by giving its projections onto the different directions of $\mathcal{K}_0$. The full map is then given by polarization. 
We consider the charge conjugation $C$ in eleven dimensions. It is skew-symmetric and obeys 
\[
C(\xi,\eta)=C(\xi_1,\eta_2)-C(\eta_1,\xi_2)\,.
\]
We write
\begin{equation}\label{decomp}
\begin{aligned}
\{\xi,\xi\}= & \{\xi,\xi\}^+K_{(+)}+\{\xi,\xi\}^-K_{(-)}+\sum\limits_{i=1}^9\{\xi,\xi\}^iK_{(i)}\\
			&+\sum\limits_{i=1}^9\{\xi,\xi\}^{i^*}K_{(i^*)}
			 +\tfrac{1}{2}\sum\limits_{i,j=1}^9\{\xi,\xi\}^{ij}K_{(ij)} 
\end{aligned}
\end{equation}
with
\begin{equation}\label{bracket-1}
\begin{gathered}
\{\xi,\xi\}^+= C(\xi_1^0,\Gamma_-\xi_1^0)\,,\ 
\{\xi,\xi\}^-=  C(\xi_2^0,\Gamma_+\xi_2^0)\,,\ 
\{\xi,\xi\}^i=  2C(\xi_2^0,\Gamma_i\xi_2^0)\,,\\
\{\xi,\xi\}^{i^*}
 				 = \frac{2i}{\lambda_i}C(\xi_1^0,\Gamma_{123}\Gamma_i\xi_2^0)\,,\\
\{\xi,\xi\}^{ij} 
				= 
		\begin{cases}
			-i\lambda_1 C(\xi_2^0,\Gamma_{123}\Gamma_{ij}\xi_2^0) & \text{ for }i,j\in \{1,2,3\}\\
			i\lambda_2 C(\xi_2^0,\Gamma_{123}\Gamma_{ij}\xi_2^0) & \text{ for }i,j\in \{4,\ldots,9\}
		\end{cases}
\end{gathered}
\end{equation}
In fact, a calculation similar to those in \cite{Klinker2014b} proves the compatibility of $\mathcal{L}$ and $\{\cdot,\cdot\}$ and yields the following result. It is an extension of the results obtained in \cite{FigPap} and a special case of a more general classification result on supergravity backgrounds of Cahen-Wallach type of which a publication is in preparation.
\begin{prop}\label{prop-1}
The CW-space given by $B=-{\rm diag}\left((\alpha-\beta)^2\mathbbm{1}_3,(\alpha+\beta)^2\mathbbm{1}_6\right)$ together with the connection $D$ described in (\ref{conn-gen}) with $(c,d)=(\alpha\Gamma_{123},\beta\Gamma_{123})$ defines a 1-parameter family of non-restricted irreducible geometric superalgebra if the odd-odd-bracket is defined as in (\ref{bracket-1}).
\end{prop}

\subsection{A family of non-restricted $N=1$, $\nu=\sfrac{3}{4}$ geometric superalgebras}

We follow the way of construction from the preceding section. For this we consider now a connection that is defined by a pair of Clifford elements
\begin{equation}\label{conn:2}\begin{aligned}
	c& =(\alpha\Gamma_{12}+\beta\Gamma_{45})\Gamma_3 = (\alpha_+X_{1245}^++\alpha_-X_{1245}^-)\Gamma_{123} \\
	d& =(\alpha'\Gamma_{12}+\beta'\Gamma_{45})\Gamma_3 = (\alpha'_+X_{1245}^++\alpha'_-X_{1245}^-)\Gamma_{123} 
\end{aligned}\end{equation}
where we introduced the combinations 
$\alpha_\pm=\alpha\mp\beta$, $\alpha'_\pm=\alpha'\mp\beta'$ and the projection operators 
$X_{1245}^\pm=\frac{1}{2}(\mathbbm{1}\pm\Gamma_{1245})$. 
The quadratic map $q_{c,d}$ associated to these elements is given by
\begin{equation}
\q_{c,d}(e_i)=
\begin{cases}
(\alpha_--\alpha_+')^2\Gamma_iX^+_{1245}+(\alpha_+-\alpha_-')^2\Gamma_iX^-_{1245} & \text{for }i\in\{1,2\}\,,\\
(\alpha_+-\alpha_+')^2\Gamma_iX^+_{1245}+(\alpha_--\alpha_-')^2\Gamma_iX^-_{1245} & \text{for }i\in\{3\}\,,\\
(\alpha_-+\alpha_+')^2\Gamma_iX^+_{1245}+(\alpha_++\alpha_-')^2\Gamma_iX^-_{1245} & \text{for }i\in\{4,5\}\,,\\
(\alpha_++\alpha_+')^2\Gamma_iX^+_{1245}+(\alpha_-+\alpha_-')^2\Gamma_iX^-_{1234} & \text{for }i\in\{6,7,8,9\}\,.
\end{cases}
\end{equation}
Therefore, we consider the CW-space $M_B$ with 
\begin{equation}\label{metric-2}
B=-{\rm diag}\left(
(\alpha_--\alpha_+')^2\mathbbm{1}_2,
(\alpha_+-\alpha_+')^2\mathbbm{1}_1,
(\alpha_-+\alpha_+')^2\mathbbm{1}_2,
(\alpha_++\alpha_+')^2\mathbbm{1}_4  
\right).
\end{equation}
It is indecomposable if and only if $\alpha_\pm\neq\pm\alpha'_+$.
As before, we write $\lambda_j^2=-(\alpha_--\alpha_+')^2$ for $j\in \{1,2\}$,
$\lambda_3^2=-(\alpha_+-\alpha_+')^2$, 
$\lambda_j^2=-(\alpha_-+\alpha_+')^2$ for $j\in \{4,5\}$,
$\lambda_j^2=-(\alpha_++\alpha_+')^2$ for $j\in \{6,\ldots,9\}$, and set $\lambda_j=-i(\alpha_\pm\pm\alpha_+')$.
The connection defined by $(c,d)$ above is non-flat such that $\nu=1$ cannot be obtained. 
The parallel spinors in this situation define $\mathcal{K}_1$ and are given by 
\begin{equation} 
	\mathcal{K}_1=\left\{\xi\in\cancel S\left|
	\begin{array}{l@{\,}l}
		\xi=\xi(\xi_1^0,\xi_2^0)= &  \exp(x^-c)\xi_1^0 \\
		& +\Big(1-\frac{1}{2}\sum_i\Gamma_+x^is_{c,d}(e_i)\Big)
		 	\exp(x^-d)\xi_2^0  \\[0.5ex]
		\multicolumn{2}{l}{\xi_1^0,\xi_2^0\text{ constant},\ \sigma_-\xi_2^0=\sigma_+\xi_1^0=0,\ X_{1245}^-\xi_2^0=0}
	\end{array}
	\right.\right\}
\end{equation}
such that $\nu=\sfrac{3}{4}$. 
Again we have  $\mathcal{L}_X\xi\in\mathcal{K}_1$ for all $\xi=\xi(\xi_1^0,\xi_2^0)\in\mathcal{K}_1$ with the same relations as before, namely (\ref{L-rel}).

We also complete the structure by introducing the map $\{\cdot,\cdot\}:\mathcal{K}_1\times\mathcal{K}_1\to\mathcal{K}_0$ as before. In this case it is given by 
\begin{equation}\label{bracket-2}
\begin{gathered}
\{\xi,\xi\}^+= C(\xi_1^0,\Gamma_-\xi_1^0)\,,\ 
\{\xi,\xi\}^-= C(\xi_2^0,\Gamma_+\xi_2^0)\,,\ 
\{\xi,\xi\}^i= 2C(\xi_2^0,\Gamma_i\xi_2^0)\,,\\
\{\xi,\xi\}^{i^*} 
			 = \frac{2i}{\lambda_i}C\big( \xi_1^0,  \Gamma_{125}\Gamma_i\xi_2^0\big)\,,\\
\{\xi,\xi\}^{ij} 
			= 
	\begin{cases}
		i\lambda_1  C\big(\xi^0_2,\Gamma_+\Gamma_{3}\xi_2^0\big) & \text{for } (ij)=(12)\\
		i\lambda_3  C\big(\xi^0_2,\Gamma_+\Gamma_{12345}\xi_2^0\big) & \text{for }(ij)=(45)\\
		i\lambda_6  C\big(\xi^0_2,\Gamma_+\Gamma_{123ij}\xi_2^0\big) & \text{for } i,j\in\{6,\ldots,9\}
	\end{cases}
\end{gathered}
\end{equation}
The involved calculations that prove the compatibility of $\mathcal{L}$ and $\{\cdot,\cdot\}$ can be found in \cite{Klinker2014b}. 

We summarize the above in the following statement.
\begin{prop}\label{prop-2}
	The CW-space that is given by the symmetric map
	$B=-{\rm diag}\left( 
		(\alpha_--\alpha_+')^2\mathbbm{1}_2,
		(\alpha_+-\alpha_+')^2,		
		(\alpha_--\alpha_+')^2\mathbbm{1}_2,
		(\alpha_++\alpha_+')^2\mathbbm{1}_4
	\right)$
	together with the connection $D$ described in (\ref{conn-gen}) with $(c,d)=	\big(
	(\alpha_+X_{1245}^++\alpha_-X_{1245}^-)\Gamma_{123},\alpha'_+X_{1245}^+\Gamma_{123}\big)$
	defines a 3-parameter family of non-restricted irreducible geometric superalgebras with $\nu=\sfrac{3}{4}$ if the odd-odd-bracket is defined as in (\ref{bracket-2}).
\end{prop}

\begin{rem}\label{rem:brackets}
\begin{itemize}
\item
The first three bracket projections in (\ref{bracket-1}) and (\ref{bracket-2}) are the analog of the usual supersymmetry brackets as known from the super Poincar\'e-algebra in the flat situation. In more common notation it reads as $\{Q_\alpha,Q_\beta\}^\mu=\Gamma^\mu_{\alpha\beta}$. The two further projections are strongly related to the ingredients that enter into the definition of the superalgebra, namely the coefficients of the connection that defines the odd summand.
\item
In the $N$-extended situation the charge conjugation is replaced by the tensor product of a charge conjugation on the first factor and a bilinear form on the second factor in the construction of the odd-odd bracket. 
\end{itemize}
\end{rem}

\begin{rem}\label{rem:iso}
	The parameters of the families of geometric algebras in Propositions \ref{prop-1} and \ref{prop-2} can be reduced by one if we identify isometric Cahen Wallach spaces, so we are left with a 1-parameter family and a 2-parameter family, respectively. 
\end{rem}

\subsection{Supersymmetry algebras}

Propositions \ref{prop-1} and \ref{prop-2} tell us what the geometric superalgebras look like. The next question we will discuss is: when does such algebra yield a supersymmetry algebra? Or: what CW-space can be considered as supergravity background? The obstruction to this is the cubic spinorial condition \ref{item:susy}, namely $\mathcal{L}_{\{\xi,\xi\}}\xi=0$. Using (\ref{decomp}) this is 
\begin{equation}
\begin{aligned}
\mathcal{L}_{\{\xi,\xi\}}\vec\xi
=\ & \{\xi,\xi\}^+\mathcal{L}_{K_{(+)}}\xi 
+\{\xi,\xi\}^-\mathcal{L}_{K_{(-)}}\xi 
+\sum_i\{\xi,\xi\}^i \mathcal{L}_{K_{(i)}}\xi \\
&	+\sum_i\{\xi,\xi\}^{i^*} \mathcal{L}_{K_{(i^*)}}  \xi
+\frac{1}{2}{\sum\limits_{ij}}^*\{\xi,\xi\}^{ij} \mathcal{L}_{K_{(ij)}}\xi
\end{aligned}
\end{equation}
If we split this into its two components we see that it yields one cubic equation for the constant spinor $\xi_2^0$ and one cubic equation for the two constant spinors $\xi_1^0,\xi_2 ^0$. 
For example, in case of the geometric superalgebra from Proposition \ref{prop-1} the two equations for the second and first components are
\begin{equation}\label{2-comp}
\begin{aligned}
0=\ &  
\beta  C(\xi^0_2,\Gamma_+\xi^0_2)\Gamma_{123}\xi_2 
	+\frac{1}{4}(\alpha-\beta)\!\!\!
		\sum_{ij\in \{1,2,3\}}\!\!\! C(\xi^0_2,\Gamma_{123}\Gamma_{ij}\xi^0_2) \Gamma_{ij}\xi^0_2\\
& 	-\frac{1}{4}(\alpha+\beta)\!\!\!\!
		\sum_{ij\in\{4,\ldots,9\}}\!\!\!\! C(\xi^0_2,\Gamma_{123}\Gamma_{ij}\xi^0_2) \Gamma_{ij}\xi^0_2
\end{aligned}
\end{equation}
and
\begin{align}
0=\ &
\alpha (\xi_2,\Gamma_+\xi^0_2) \Gamma_I\xi_1 \nonumber\\
&	+ (\alpha-\beta)\!\!\!\sum_{i\in \{1,2,3\}} \!\!\!
			\big( C(\xi^0_1,\Gamma_i\xi^0_2)\Gamma_{123}\Gamma_i\xi_2^0
				-C(\xi_1^0,\Gamma_{123}\Gamma_i\xi_2)\Gamma_i\xi_2^0\big)\nonumber\\	
&	+ (\alpha+\beta)\!\!\!\!\sum_{i\in\{4,\ldots,9\}}\!\!\!\! 
			\big( C(\xi^0_1,\Gamma_i\xi^0_2)\Gamma_{123}\Gamma_i\xi^0_2
				-C(\xi_1^0,\Gamma_{123}\Gamma_i\xi_2^0)\Gamma_i\xi_2^0\big) \label{1-comp}\\ 
&	+\frac{1}{4}(\alpha-\beta)\!\!\!\sum_{ij\in \{1,2,3\}}\!\!\! 
				C^(\xi^0_2,\Gamma_{123}\Gamma_{ij}\xi^0_2) \Gamma_{ij}\xi_1^0\nonumber\\
& 	-\frac{1}{4}(\alpha+\beta)\!\!\!\!\sum_{ij\in\{4,\ldots,9\}}\!\!\!\!
				C^(\xi^0_2,\Gamma_{123}\Gamma_{ij}\xi^0_2) \Gamma_{ij}\xi_1^0\,. \nonumber
\end{align}
The only combinations of coefficients for which (\ref{1-comp}) and (\ref{2-comp}) can be identically solved is $\alpha=-3\beta$.

The analogue of (\ref{2-comp})-(\ref{1-comp}) for the superalgebras from Proposition \ref{prop-2} can be found in \cite{Klinker2014b} and the only choice of coefficients that solve the resulting equations is $\alpha_+=-3\alpha_+'$.

We collect the results in the following proposition.
\begin{prop}\label{prop:susy}
\begin{enumerate}
	\item \label{item:1} The geometric superalgebra according to Proposition \ref{prop-1} yields non restricted $\nu=1$ geometric supersymmetry if and only if 
	\[
	B=-4\alpha^2{\rm diag}\left(
	4\mathbbm{1}_3,
	\mathbbm{1}_6  
	\right)
	\]
	and $(c,d)$ is given by
	\[
	c = -3\alpha\Gamma_{123},\  
	d = \alpha\Gamma_{123}\,. 
	\]
	\item \label{item:2} The geometric superalgebra according to Theorem \ref{prop-2} yields non restricted $\nu=\sfrac{3}{4}$ geometric supersymmetry if and only if 
	\[
	B=-{\rm diag}\left(
	(\alpha_--\alpha_+')^2\mathbbm{1}_2,
	16\alpha_+'^2\mathbbm{1}_1,	
	(\alpha_-+\alpha_+')^2\mathbbm{1}_2,
	4\alpha_+'^2\mathbbm{1}_4  
	\right)
	\]
	and $(c,d)$ is given by
	\[
	c = \big(-3\alpha_+' X^+_{1245}+\alpha_- X^-_{1245}\big)\Gamma_{123},\  
	d = \alpha_+' X^+_{1245}\Gamma_{123}\,. 
	\]
\end{enumerate}
\end{prop}
\begin{rem}\label{rem:iso2}
Part one of the above Proposition is exactly the unique maximal supergravity background of CW-type that has been considered in \cite{CKG,FigPap}. The uniqueness follows after identifying isometric spaces, see Remark \ref{rem:iso}.
\end{rem}

\section{A common reduction}

Consider two CW-spaces of dimension $D'=n'+2$ and $D=n+2>D'$ associated to the symmetric maps $B'$ and $B=B'\oplus B''$. These two spaces come with geometric superalgebras $\mathcal{K}'$ and $\mathcal{K}$ that are $N'$-extended and $N$-extended. Furthermore the relation $N'=2^{\big[\frac{n-n'}{2}\big]}N$ holds.

Then $\mathcal{K'}$ is called a reduction of $\mathcal{K}$ -- $\mathcal{K}$ an oxidation of $\mathcal{K}'$ -- if the following holds: 
We erase from $\mathcal{K}_0$ exactly $n-n'$ Killing vector fields from $K_{(i)}$ and the same from $K_{(i^*)}$ such that the remaining part $\mathcal{K}$ remains an algebra and is isomorphic to $\mathcal{K}'$ -- maybe after restricting $\mathcal{K}_1$. This is in particular of interest in the following situations:
\begin{itemize}
\item
Can we reduce and/or oxidate a geometric supersymmetry so that the result is such, too?
\item
Suppose no member of a family of geometric superalgebras yields geometric supersymmetry, can we find a reduction of a member that does? 
\end{itemize}
We will address these question with regard to the families we presented before.

\subsection{The first reduction}

Due to the nature of the connection defined by (\ref{conn:2}) the odd part of the super Lie algebra from Proposition \ref{prop:susy}-\ref{item:2} cannot be restricted further in the generic situation. Nevertheless, there is one configuration of parameters where this is possible, namely $\alpha_+'=0$. In this situation, we are left with a decomposable CW-space associated to $B=-\alpha^2{\rm diag}(1,1,0,1,1,0,0,0,0,0)$ and the odd part of the geometric supersymmetry is restricted to $X_{1245}^+\xi_1^0=0$ in addition to $X_{1245}^-\xi_2^0=0$, i.e.\ $\nu=\frac{1}{2}$.

By taking a closer look at (\ref{2-comp})-(\ref{1-comp}) we see that both terms vanish for a truncated summation over $\{1,2,4,5\}$ if the proposed restriction is performed. The resulting algebra can then be interpreted as a reduced $4$-extended geometric supersymmetry in the following way. These are exactly the data for the $D=6$, $N=4$ supergravity background proposed in \cite{Meessen2002}.

Consider the six-dimensional CW-space associated to $B=-\alpha^2\mathbbm{1}_4$ and its spinor bundle $S_6$ with charge conjugation $C_6$. Within $S=S_6\otimes \CC^4$ we identify the second factor with the five-dimensional spin-representation and provide it with the charge conjugation $C_5$. Then $C=C_6\otimes C_5$ defines a bilinear form on $S$.  
Furthermore we consider the connection defined by 
\begin{equation}
c=\alpha X_{1234}^- \Gamma_{12}\otimes T ,\ d=0\,.
\end{equation}
Here $T$ is some vector in $\RR^5$ regarded as an element of the Clifford algebra with $T^2=-\mathbbm{1}$. 
The parallel spinors of this connection are parametrized by constant spinors $\xi_1^0\otimes v_1\in S_{6,-}\otimes\CC^5,\xi_2^0\otimes v_2\in X_{1234}^+S_{6,+}\otimes\CC^5$.

To define $\mathcal{K}_1$ we impose the further condition $X_{1234}^+\xi_1^0\otimes v_1=0$ that yields the reduction to $\nu=\sfrac{1}{2}$.

We use $C$ to write down the bracket structure of the algebra: 
\begin{equation}\label{brackets}
\begin{aligned}
\{\xi,\eta\}^{+} &= C_6(\xi_1^0,\Gamma_-\eta_1^0)C_5(v_1,w_1)\,,\\	
\{\xi,\eta\}^{-} &= C_6(\xi_2^0,\Gamma_+\eta_2^0)C_5(v_2,w_2)\,,\\	
\{\xi,\eta\}^{i}\, &= C_6(\xi_1^0,\Gamma_i\eta_2^0)C_5(v_1,w_2)+C_6(\eta_1^0,\Gamma_i\xi_2^0)C_5(w_1,v_2)\,,\\
\{\xi,\eta\}^{i^*}\!\!	
&= \frac{1}{\alpha}\big(C_6(\xi_1^0,\Gamma_{12}\Gamma_i\eta_2^0)C_5(v_1,Tw_2)
+C_6(\eta_1^0,\Gamma_{12}\Gamma_i\xi_2^0)C_5(w_1,Tv_2)\big),
\\		
\{\xi,\eta\}^{ij}\!
&= - \alpha C_6\big(\xi^0_2,\Gamma_+\eta^0_2)C_5(v_2,Tw_2)
\end{aligned}
\end{equation}
for $\xi=\xi(\xi_1^0\otimes v_1,\xi_2^0\otimes v_2)$, $\eta=\eta(\eta_1^0\otimes w_1,\eta_2^0\otimes w_2)$.

\begin{prop}\label{prop:firstred}
The above data yield a six dimensional $4$-extended non-restricted geometric supersymmetry with $\nu=\sfrac{1}{2}$ that is the reduction of the eleven-dimensional geometric supersymmetry from Proposition \ref{prop:susy}-\ref{item:2} for $\alpha_+'=0$.
\end{prop}

\subsection{The second reduction}

As we know, the non-restricted geometric superalgebras with $\nu=1$ from Proposition \ref{prop-1} only yield geometric supersymmetry for a special choice of coefficients, see Proposition \ref{prop:susy}-\ref{item:1}. Nevertheless, if we again take a look at (\ref{2-comp})-(\ref{1-comp}) for the choice $\beta=0$ we see that the truncation to $\{1,2,4,5\}$ annihilates both sums if we consider a restriction to $\nu=\sfrac{1}{2}$ of the odd part that is similar to the one before, namely $X_{1245}^+\xi_1^0=X_{1245}^-\xi_2^0=0$. The interpretation of this truncation as a reduction is as follows

We consider the same six-dimensional CW-space and the same spinor bundle as above but with connection defined by
\begin{equation}
c=\alpha\Gamma_{12}\otimes T,\ d=0\,.
\end{equation}
This connection is flat and the parallel spinors are parametrized by all of $\xi_1^0\otimes v_1\in S_{6,-}\otimes \CC^4,\xi_2^0\otimes v_2\in  S_{6,+}\otimes \CC^4$. We consider the restriction subject to the conditions $X_{1234}^+\xi_1^0=X_{1234}^-\xi_2^0=0$ as before.
Furthermore, we use the same brackets as in (\ref{brackets}).

\begin{prop}\label{prop:secondred}
	The six-dimensional data above yield a restricted 4-extended geometric supersymmetry with $\nu=\sfrac{1}{2}$ that is a reduction of the eleven-dimensional non-restricted geometric superalgebra from Proposition \ref{prop-1} for $\beta=0$. 
\end{prop}

\subsection{Concluding remarks}

\begin{itemize}
	\item
	The correspondences that have been claimed in Propositions \ref{prop:firstred} and \ref{prop:secondred}  can be made precise by identifying $T$ with $\Gamma_{3}$ and embedding $M_6$ into $M_{11}$ by $(\pm,1,2,3,4)\to (\pm,1,2,4,5)$.
	\item
	We will briefly explain, why the two six-dimensional supersymmetries constructed in the last to subsections are essentially the same although the connections that define the structures are not.\\	
	In the construction of both supersymmetries we could have forget about the further restriction of $\xi_1^0$. That would also lead to supersymmetries and to six-dimensional supergravity backgrounds but with $\nu=\sfrac{3}{4}$ in both cases. The first one would then be non-restricted and the second one would be restricted, by definition.\\
	However, by introducing the further condition on $\xi_1^0$ we guarantee that the algebra structures in both cases coincide. Roughly, this due to the fact that both connections differ by a half spinor projection that enters into the bracket structure (\ref{brackets}). In case of additional restriction this is only artificially present so that we could omit it.
	\item
	We want to emphasize the differences of the two oxidation of Propositions \ref{prop:firstred} and \ref{prop:secondred}. \\
	In the first case the eleven-dimensional CW space is decomposable with non-flat connection and the geometric superalgebra is indeed a super Lie algebra. Therefore, oxidation and reduction can both be considered as supergravity backgrounds.\\ 
	In the second case the oxidation is an indecomposable CW-space with flat connection but the superalgebra does not define supersymmetry. Nevertheless the space belongs to a family of geometric superalgebras that contains a supersymmetric solution, namely the sole maximal supergravity background of CW-type, see Remark \ref{rem:iso2}
\end{itemize}


\begin{thebibliography}{99}
	
\bibitem{CKG}
Piotr T.~Chru\'{s}ciel and Jerzy Kowalski-Glikman (1984)
	\newblock {\em Phys.\ Lett.\ B} {\bf 149} 107-110.

\bibitem{FigPap}
Jos{\'e} Figueroa-O'{}Farrill and George Papadopoulos (2001)
	\newblock {\em J.~High Energy Phys.} 2001 {\bf 08} 036.

\bibitem{GauntHull}
Gerome P.~Gauntlett and Christopher M.~Hull (2002) 
	\newblock {\em J. High Energy Phys.} 2002 {\bf 06} 013.

\bibitem{Fig03}
Jos{\'e} Figueroa-O'{}Farrill (2003)
	\newblock {\em Class.\ Quant.\ Grav.} {\bf 20} 3327-3340.

\bibitem{MeFig04}
Jos{\'e} Figueroa-O'{}Farrill, Patrick Meessen, and Simon Philip (2005)
	\newblock {\em Class.\ Quant.\ Grav.} {\bf 22} 207-226.

\bibitem{Hustler}
Jos\'{e} Figueroa-O'Farrill and  Noel Hustler (2012) 
	\newblock {\em Class.\ Quant.\ Grav.} {\bf 30} 045008.

\bibitem{CW}
Michel Cahen and Nolan Wallach (1970) 
	\newblock {\em Bull.\ Amer.\ Math.\ Soc.} {\bf 76} 585-591.

\bibitem{Wu}
Hung-Hsi Wu (1967)
	\newblock {\em Pacific J.\ Math.} {\bf 20} no.~2 351-392.
	
\bibitem{Klinker2014a}
Frank Klinker (2014) 
	\newblock {\em Adv.\ Appl.\ Clifford Algebr.} 32 pp  
	\newblock \href{http://dx.doi.org/10.1007/s00006-014-0451-7}{doi:10.1007/s00006-014-0451-7}.

\bibitem{Kosman}
Yvette Kosmann (1971)
	\newblock {\em Ann.\ Mat.\ Pura Appl.} {\bf 91} no.~1 317-395.

\bibitem{Klinker2014b}
Frank Klinker (2014)
	\newblock  {\em submitted}, Preprint: \href{http://arxiv.org/abs/1406.4672}{arXiv:1406.4672 [math.DG]}.

				
\bibitem{Meessen2002}
Patrick Meessen (2002)
	\newblock {\em Phys.\ Rev.\ D} {\bf 65} 087501.



\end{thebibliography}
\end{document}